\documentclass[twocolumn,showpacs,superscriptaddress,preprintnumbers,amsmath,amssymb,prc]{revtex4-1}
\usepackage{amsmath}
\usepackage{cases}
\usepackage{amssymb}
\usepackage{CJK}% Chinese character support
\usepackage{graphicx}% Include figure files
\usepackage{dcolumn}% Align table columns on decimal point
\usepackage{bm}% bold math
\usepackage{color}
\usepackage{ulem}%lines on text
\usepackage[pagewise]{lineno}%ÌíŒÓÐкÅ
\begin{document}
\begin{CJK*}{GB}{gbsn}

\title{Azimuthal anisotropy and multiplicities of hard photons and free nucleons in intermediate-energy heavy-ion collisions}

\author{S. S. Wang (ÍõÉÁÉÁ)}
\affiliation{Key Laboratory of Nuclear Physics and Ion-beam Application (MOE), Institute of Modern Physics, Fudan University, Shanghai 200433, China}
\affiliation{Shanghai Institute of Applied Physics, Chinese Academy of Sciences, Shanghai 201800, China}
\affiliation{University of Chinese Academy of Sciences, Beijing 100049, China}

\author{Y. G. Ma (ÂíÓà¸Õ)}\thanks{Email: Corresponding author. mayugang@fudan.edu.cn}
\affiliation{Key Laboratory of Nuclear Physics and Ion-beam Application (MOE), Institute of Modern Physics, Fudan University, Shanghai 200433, China}

\author{X. G. Cao (²Üϲ¹â)}
\affiliation{Shanghai Advanced Research Institute, Chinese Academy of Sciences, Shanghai 201210, China}

\author{D. Q. Fang (·½µÂÇå)}
\affiliation{Key Laboratory of Nuclear Physics and Ion-beam Application (MOE), Institute of Modern Physics, Fudan University, Shanghai 200433, China}

\author{C. W. Ma (Âí´ºÍú)}
\affiliation{Institute of Particle and Nuclear Physics, Henan Normal University, Xinxiang 453007, China}
\date{\today}

\begin{abstract}
Anisotropic flow can offer significant information of evolution dynamics in heavy-ion collisions. A systematic study of the directed flow $v_1$ and elliptic flow $v_2$ of hard photons and free nucleons is performed for $^{40}$Ca+$^{40}$Ca collisions in a framework of isospin dependent quantum molecular dynamics (IQMD) model. The study firstly reveals that thermal photons emitted in intermediate-energy heavy-ion collisions have the behaviors of directed and elliptic flows. The interesting phenomena of incident energy dependence of  $v_1$ and  $v_2$  for thermal photons in central collisions also confirmed that it can be regarded as a good probe of evolution dynamics. Moreover, the multiplicities of hard photons and free nucleons and their correlation are also investigated. We find that direct photon emission is positively related to free nucleons emission, however,  there exists an anti-correlation  for thermal photons with  free nucleons.
\end{abstract}

\maketitle
\section{Introduction}
\label{introduction}
Heavy-ion collisions make it possible to investigate on the properties of nuclear matter, especially the phase diagram of nuclear matter where a transition from the Fermi liquid ground state to the nucleon gas phase  was predicted \cite{Bord,MaCW,Ma_onset,GSI,MaPRL,Natowitz,Ma_data1,Ma_data2,Ma_NST,LiuHL}. Due to the considerable advantage of hard photons not being disturbed by the final-state interactions except for weakly interacting with the surrounding nuclear medium through the electromagnetic interaction, they are very clean probes of the reaction dynamics and can deliver an undisturbed picture of the emitting source \cite{YS1998,Bona,Cassing,YGM2012,GHL2008,Deng2018magnetic,Yong,Nif}. Therefore, particular attentions are paid to hard photons emitted from intermediate-energy heavy-ion collisions since hard photons were observed in experiments \cite{KBB1985}. So far experimental \cite{CLT1988,MK1988,JS1986,EG1986,RB1986,EM1993} and theoretical works \cite{BAR1986,BAR1987,CMK1985,CMK1986,KN1986,KN1988,TSB1987,WB1986,WC1986,DTK1991,GM1995,FM1995,zhang2019time,bonasera1992kinetic} demonstrated that hard photons mainly originate from incoherent bremsstrahlung from individual neutron-proton collisions, which own two distinct sources (i.e. direct photon and thermal photon sources) in space and time according to the experimental evidence as well as the Boltzmann-Uehling-Uhlenbeck (BUU) model calculations \cite{FM1995,GM1995}. Although there are some hot fragments formed in heavy-ion collisions which could result in a larger width of giant dipole resonance (GDR) energy spectrum and the energy of GDR photons might be higher than 30MeV \cite{nuovo2000}, the GDR strength is much weaker than the bremsstrahlung photon cross section. So generally, the photons with energy above 30MeV is dominated by hard photons \cite{Schutz1997}. Meanwhile, extreme high energy hard photon as a unique probe for the properties of quark-gluon plasma has also been extensively discussed in ultra-relativistic heavy-ion collision community \cite{Phenix,Gale,Roy,Heinz,Rapp,LongJL,HeZJ,Dasgupta,Golo}.

Azimuthal anisotropy as a very useful probe has been widely used to explore the hot dense matter in relativistic heavy-ion collisions, especially the directed flow $v_1$ and elliptic flow $v_2$ \cite{Phenix,Gale}. Recently, it was confirmed that direct photons emitted from intermediate-energy heavy-ion collisions also show the behaviors of $v_1$ and $v_2$ in the framework of BUU model~\cite{YGM2012,GHL2008,Deng2018magnetic}. But for thermal photons, it is still unknown whether they have the behaviors of azimuthal asymmetry. Considering that hard photons production happens throughout the whole dynamical evolution of nuclear reaction, $v_1$ and $v_2$ are expected to be a powerful observable of reaction dynamics in intermediate-energy heavy-ion collision. In this context, the hard photon production channel, i.e. $n +p \rightarrow n + p + \gamma$, is embedded into an isospin dependent quantum molecular dynamics (IQMD) model taking into account the in-medium effect. A systematic study of azimuthal anisotropy for direct photons, thermal photons and free nucleons will be performed in this version of IQMD model. By comparing between them, we expect to learn some information of reaction dynamics.

The paper is organized as follows: In Sec. \ref{modelformalism}, we give a brief description of the IQMD model and introduce  the formula of hard photon production probability, and the definition of direct photons and thermal photons as well as the definition of anisotropic flows. Results and discussion are described in Sec. \ref{resultsanddiscussion}, including the collision centrality and incident energy dependence of directed flows and elliptic flows for hard photons and free nucleons, respectively. In addition, we also demonstrate the characteristics of multiplicity of hard photons and free nucleons as well as the correlation between them. Finally, a summary is given  in Sec.~\ref{summary}.

\section{MODEL AND FORMALISM}
\label{modelformalism}
\subsection{Brief introduction of the IQMD model}

The isospin-dependent quantum molecular dynamics model was developed from the standard QMD model, which has been successfully applied to describe the intermediate-energy heavy-ion collisions \cite{Aichelin,Hartnack,qmd,Yan1,WangTT,Yan2,LiPC,Ono,ZhangZF,FengZQ,Sood}. In this model, each nucleon is represented by a Gaussian wave function with a width $L$ (here we take $L$ = 2.16 $fm^2$) as \cite{Aichelin}
\begin{equation}\label{wavesingle}
\phi_{i}(\textbf{r},t) = \frac{1}{(2\pi L)^{3/4}} exp[ -\frac{(\textbf{r}-\textbf{R}_{i})^{2}}{4 L}+\frac{i\textbf{P}_{i}\cdot \textbf{r}}{\hbar}],
\end{equation}
here $ \textbf{R}_{i} $ and $ \textbf{P}_{i} $ are the centers of position and momentum of the $i$-th wave packet, respectively. The total wave function for the $ N $-body system  is treated as a direct product of these wave function,
\begin{equation}\label{wavetotal}
\Phi(\textbf{r},t) = \prod^{N}_{i}\phi_{i}(\textbf{r},t).
\end{equation}
More detailed description can be found in Ref.~\cite{Aichelin,Hartnack,qmd}.

Considering that in-medium effects cannot be ignored in intermediate-energy heavy-ion collisions \cite{OL2014NNCS}, the screened nucleon-nucleon cross section is embedded into the model as the in-medium nucleon-nucleon cross section (in-medium NNCS) instead of the free nucleon-nucleon cross section (free NNCS) parameterized from experimental measurements \cite{KZ1968NNCS}. The formula is derived from the geometric reasoning that the geometric cross-section radius should not exceed the inter-particle distance and is implemented in the form \cite{PD2002NNCS,DW2011NNCS,OL2014NNCS}
\begin{align}\label{inmediumnncs}
\sigma^{in-medium}_{NN} = &\sigma_{0}\tanh(\sigma^{free}_{NN}/\sigma_{0}),  \\
\sigma_{0} = &y\rho^{-2/3}, ~~~y = 0.85.
\end{align}
Here $\rho$ denotes the single-particle density. In our recent work ~\cite{shan2020}, a comparison between experimental data and calculated photon energy spectra has been performed, which indicates that the calculated results employing the in-medium NNCS in the IQMD model is in a good agreement with the experimental data.

\subsection{Hard photon production probability}

The main mechanism of hard photon production in intermediate-energy heavy-ion collisions is incoherent bremsstrahlung from individual $n-p$ collisions \cite{WB1986,WC1986,CMK1985,KN1988,TSB1987}. The elementary double differential production probability in the nucleon-nucleon center-of-mass frame is described by employing the hard-sphere collision limit from Ref.~\cite{JDJ1962} and modified as in Ref.~\cite{WC1986} for energy conservation as
\begin{equation}\label{productionrate}
\frac{d^{2}P'}{dE'_{\gamma}d\Omega'_{\gamma}} = \frac{\alpha_c}{12\pi^2}\frac{1}{E'_{\gamma}}(2\beta^{2}_{f}+3sin^{2}\theta_{\gamma}\beta^{2}_{i}),
\end{equation}
where $\alpha_c$ is the fine structure constant, $ E'_{\gamma} $ is the energy of emitting photon. $ \beta_{i} $ and  $ \beta_{f} $ are the initial and final velocity of the proton, and $ \theta_{\gamma} $ is the angle between the momenta of the incident particle and the emitted photon. The double differential production probability in the nucleus-nucleus center-of-mass frame can be derived from Eq.(\ref{productionrate})
\begin{equation}\label{totprobabality}
\begin{aligned}
\frac{d^{2}P}{dE_{\gamma}d\Omega_{\gamma}} = &\sum_{pn coll}\int
\frac{d\Omega_{e}}{4\pi}\frac{E_{\gamma}}{E^{'}_{\gamma}}\frac{d^{2}P'}{dE^{'}_{\gamma}d\Omega^{'}_{\gamma}}(\textbf{k}_{1}-\textbf{k}_{2})\\
&\times[1-P_{block}],
\end{aligned}
\end{equation}
where the last term represents the effects of Pauli blocking in the final state phase space. In generally, QMD model underestimates the blocking probability because of fluctuations \cite{zhang2018Comparison}. Considering that the procedure of Pauli blocking is very important, we have checked and confirmed that the Pauli blocking effect in this IQMD model is reasonable in our recent article \cite{shan2020}. 
\subsection{Definitions of directs photon and thermal photons}

Up to now, it is well established that hard photons are emitted from two distinct sources in space and time \cite{FM1995,GM1995,GHL2008,YGM2012,zhang2019time}. Direct photons originate from dense source during the first compression-phase at the earlier stage of reaction which account for the dominant contribution, while the thermal photons are produced from a thermalized source at the later stage of reaction. Based on this definition, direct photons or thermal photons can be identified by the time evolution of system average density $\langle \rho \rangle$ \cite{GHL2008,YGM2012}. In this work, we also employ the same method which means the separation time ($t_s$) between them corresponds to the first minimum average density.

\begin{figure}
    \includegraphics[width=8.6cm]{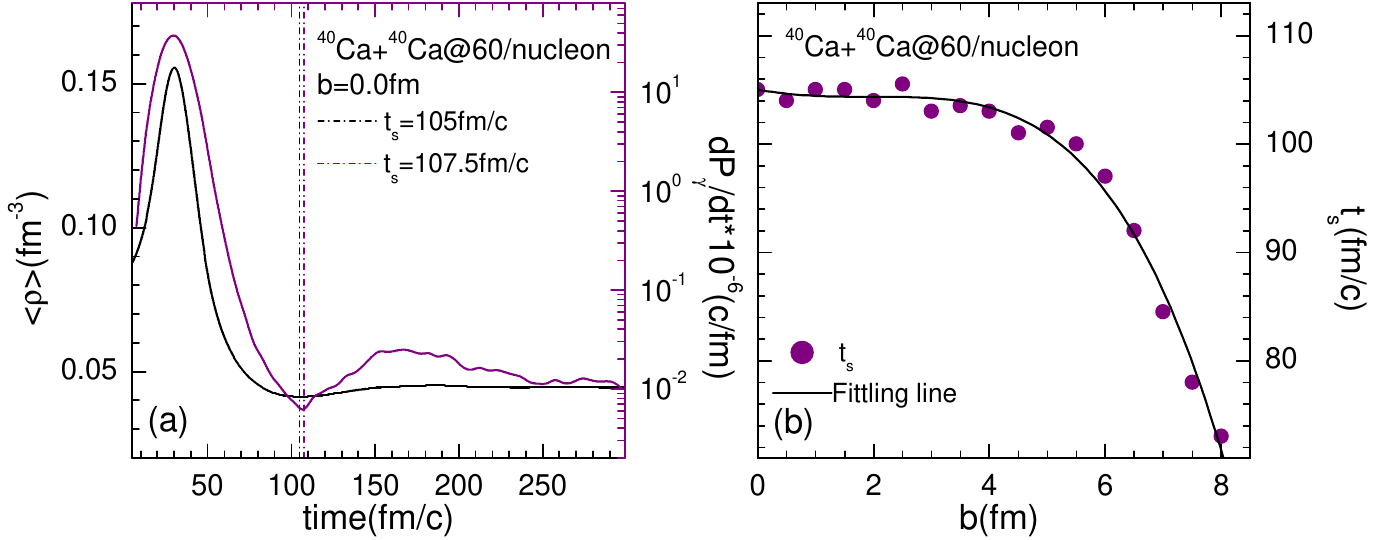}
    \caption{(a) Time evolution of average density (black line) and hard photon production probability (purple line), respectively. Dash dot lines correspond to the moments reaching their own first minimum values. (b) Separation time $t_s$ between direct photons and thermal photons as a function of impact parameters. Line is the fitting curve with a polynomial function.}
    \label{fts}
\end{figure}

As an example, Fig.~\ref{fts}(a) shows the time evolution of average density $\langle\rho\rangle$ and differential hard photon production probability in the reaction of $^{40}$Ca+$^{40}$Ca@$60$ MeV/nucleon at an impact parameter $b=0.0$ fm, respectively. It is found that the hard photon production probability is sensitive to the density during the whole evolution process. Both of them undergo a second rise tendency after the first rise and fall with time evolution. It means when the system is at the compression stage both density and photon production probability rapidly increase, and then with the system expansion their values start to decrease. After that, the system undergoes a second compression stage which makes their values increase again. The moments corresponding to their minimum values are almost consistent, i.e. $105$ fm/c for $\langle\rho\rangle$ and $107.5$ fm/c for photon production probability. It is well known that impact parameter and incident energy $E_{int}/A$ are two important factors to determine $t_s$ for the reaction system. Therefore, in order to obtain $t_s$ for every event, we firstly obtain the impact parameter dependence of $t_s$ at a given $E_{int}/A$ and then get a fitting function. For example, Fig.~\ref{fts}(b) plots $t_s$ versus impact parameter in the reaction of $^{40}$Ca+$^{40}$Ca@$60$ MeV/nucleon, where the line represents the fitting curve with a polynomial function up to 4-th order. If such a fitting function is determined, the $t_s$ at each impact parameter b for a given $E_{int}/A$ is obtained and then we can use this value to distinguish the thermal photons from the direct photons.

\subsection{Definition of anisotropic flows}

Anisotropic flow is an important observable, which reflects the azimuthal asymmetry in particle distribution with respect to the reaction plane (the plane spanned by the beam direction and the impact parameter axis). It is commonly defined as different $n-$th harmonic coefficients $v_n$ of an azimuthal Fourier  expansion of the particle invariant distribution \cite{GHL2008,YGM2012, Ma_1993,voloshin2003anisotropic, Ma2007scaling,Deng2018magnetic, voloshin1996flow},
\begin{equation}\label{anisotropicflow}
\frac{dN}{d\phi}\propto1 + 2\sum^{\infty}_{n=1}v_n cos(n\phi),
\end{equation}
where $\phi$ is the azimuthal angle between the transverse momentum of the particle and the reaction plane. The directed flow $v_1$ and elliptic flow $v_2$ correspond to the first and the second harmonic Fourier coefficient in Eq.(\ref{anisotropicflow}), respectively. The formulas are read as
\begin{gather}
\label{v1flow}
v_1 = \langle cos\phi \rangle=\langle\frac{p_x}{p_T}\rangle, \\
\label{v2flow}
v_2 = \langle cos(2\phi) \rangle=\langle\frac{p^2_x-p^2_y}{p^2_T}\rangle,
\end{gather}
where $p_T = \sqrt{p^2_x+p^2_y}$. The elliptic flow characterizes the eccentricity of the particle distribution in momentum space.

\section{RESULTS AND DISCUSSION}
\label{resultsanddiscussion}

In this paper, the reactions of $^{40}$Ca+$^{40}$Ca at incident energies $E_{int}/A=40$MeV-$120$MeV were simulated by IQMD model which takes into account  the in-medium NNCS in the process of two-body collisions. A soft nuclear equation of state is used in the present simulation.
Recently, it has been reported that direct photon emission is azimuthal anisotropic in intermediate-energy heavy-ion collisions \cite{GHL2008,YGM2012,Deng2018magnetic}. But for thermal photons, there is still barren. In addition, hard photons mainly originate from individual proton-neutron collision, and free nucleons are emitted from nucleon-nucleon collision throughout the dynamical evolution of the system. Therefore, more information of dynamical properties are expected by investigating on their azimuthal anisotropy.

\subsection{Directed flow of hard photons and free nucleons}

\begin{figure}
    \includegraphics[width=8.6cm]{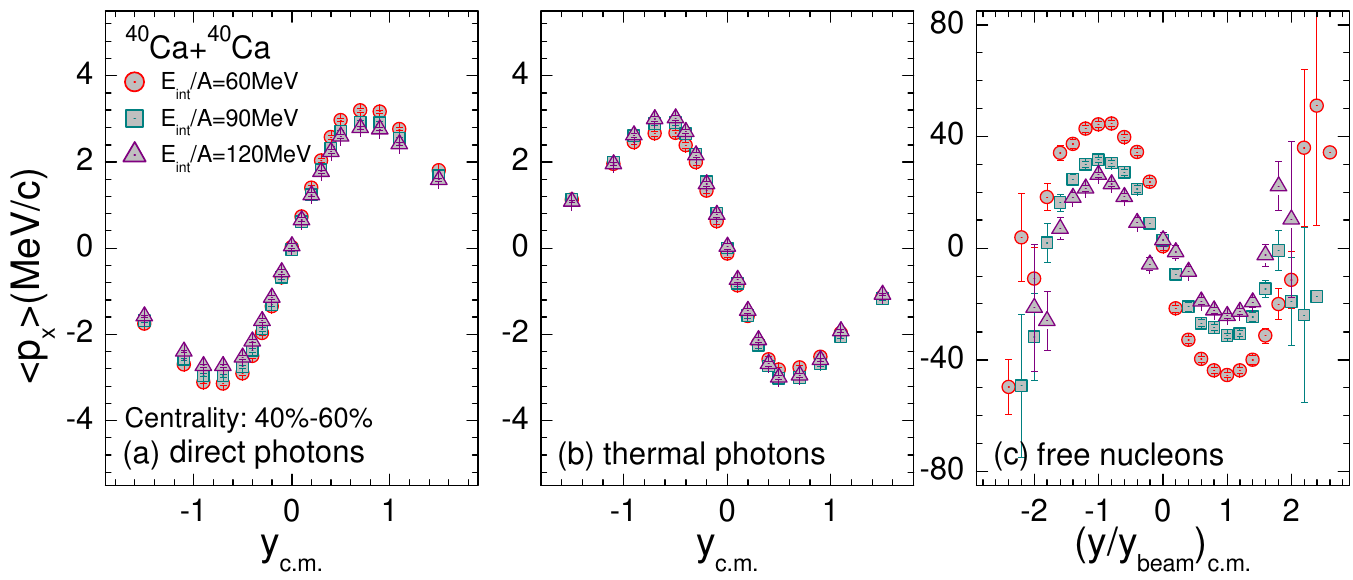}
    \caption{Average in-plane transverse momentum of (a) direct photons, (b) thermal photons and (c) free nucleons as a function of c.m. rapidity for $^{40}$Ca+$^{40}$Ca collisions at $60$MeV/nucleon (circles), $90$MeV/nucleon (squares) and $120$MeV/nucleon (triangles) in the centrality of $40\%-60\%$, respectively.}
    \label{fpxy}
\end{figure}

In this section, we focus on the investigation of directed flow of hard photons and free nucleons emitted in intermediate-energy heavy-ion collisions. Firstly,  Fig.~\ref{fpxy}(a) and (b) plot  the average in-plane transverse momentum $\langle p_{x}\rangle$ of direct photons and thermal photons as a function of the rapidity of particles in the center-of-mass frame $y_{c.m.}$ for the reaction of $^{40}$Ca + $^{40}$Ca at $60$, $90$ and $120$MeV/nucleon in the centrality of $40\%-60\%$, respectively. Note that collision centrality is defined  by $ \frac{100\pi b^2}{\pi b^2_{max}}$ where $b$ denotes impact parameter and $b_{max}$ is the summation of the radius of reaction system.
With this definition, smaller centrality corresponds to more central collisions, and larger centrality means more peripheral collisions.
Fig.~\ref{fpxy}(c) corresponds to the $\langle p_{x}\rangle$ for free nucleons versus the reduced c.m. rapidity $(y/y_{beam})_{c.m.}$. One can see that the $\langle p_{x}\rangle$ distribution with rapidity for thermal photons is similar to that for free nucleons, while it is in opposite trend with that of direct photons. Moreover, the $\langle p_{x}\rangle$ for the direct photons and thermal photons are insensitive to incident energy $E_{int}/A$. However, $E_{int}/A$ plays an important role for free nucleons. In addition, it also be found that all of them show the anti-symmetric behavior about mid-rapidity so that the following calculations for directed flow $v_1$ just need to take into account that at the positive rapidity range, especially for thermal photons which demonstrates thermal photons also have the behavior of directed flow.

\begin{figure}
    \includegraphics[width=8.6cm]{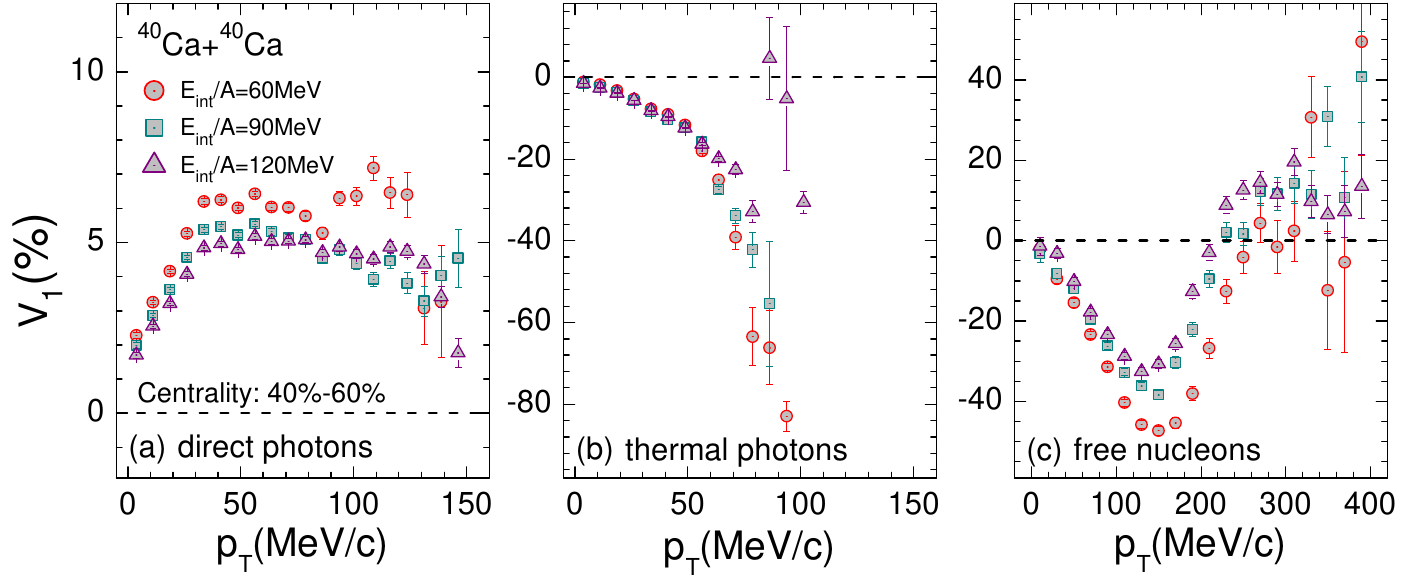}
    \caption{Directed flows of (a) direct photons, (b) thermal photons and (c) free nucleons as a function of transverse momentum for  $^{40}$Ca+$^{40}$Ca collisions at $60$MeV/nucleon (circles), $90$MeV/nucleon (squares) and $120$MeV/nucleon (triangles) in the centrality of $40\%-60\%$, respectively.}
    \label{fv1pT}
\end{figure}

Fig.~\ref{fv1pT} plots transverse momentum $p_T$ dependence of $v_1$ for direct photons, thermal photons and free nucleons in the reaction of $^{40}$Ca+$^{40}$Ca at $60$, $90$ and $120$MeV/nucleon in a centrality of $40\%-60\%$, respectively.  For direct photons, they have a rapid rise and then tend to a saturate state at $p_T >$ 40 MeV/c. The positive $v_1$  reveals that direct photons essentially stem from the repulsive mean field. Conversely, it can be seen from Fig.\ref{fv1pT} (b) and (c) that both the thermal photons and free nucleons own the negative $v_1$ which indicates they are dominated  by the attractive mean field.

\begin{figure}
    \includegraphics[width=8.6cm]{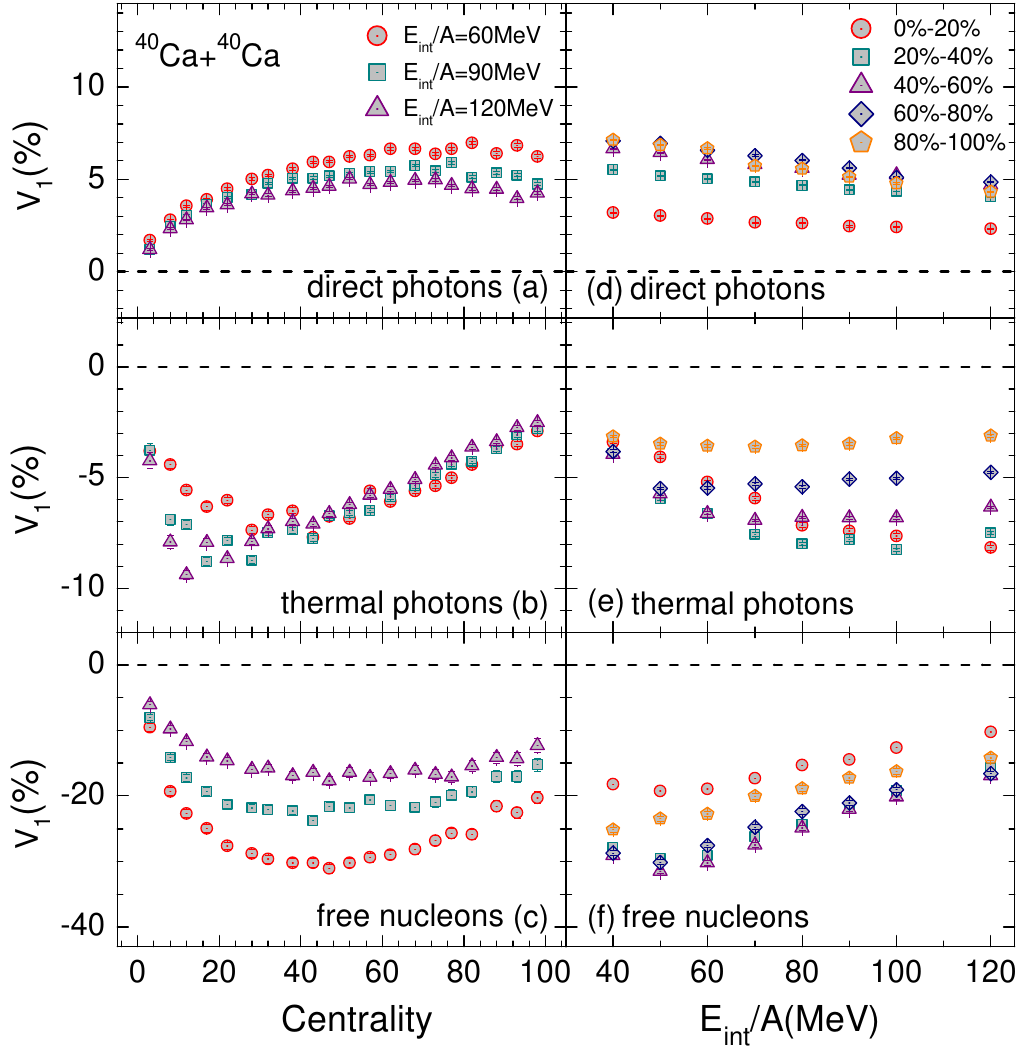}
    \caption{Directed flows of direct photons (a and d), thermal photons (b and e) and free nucleons (c and f) as a function of centrality (a-c) and incident energy (d-f) for $^{40}$Ca + $^{40}$Ca collisions, respectively. }
    \label{fv1CentrEin}
\end{figure}

Finally, the directed flows $v_1$ of direct photons, thermal photons and free nucleons as functions of collision centrality and incident energy $E_{int}/A$ are plotted in Fig.\ref{fv1CentrEin} for the reaction of $^{40}$Ca+$^{40}$Ca, respectively. From Fig.~\ref{fv1CentrEin}(a-c), it is seen that all of the absolute values of $v_1$ have a rise and then fall with centrality. This phenomenon can be understood that it is azimuthal symmetric for particle emission in central collisions, and then the effect of azimuthal asymmetry gradually enhances in mid-peripheral collisions, while with the continuous increasing of centrality, the size of reaction zone starts to decrease, which results in the diminution of azimuthal asymmetry effect. Except that, the sign of $v_1$ for direct photons is different from that of thermal photons and free nucleons which also means they are dominated by a different mean field role. By comparing the value of $v_1$,  it is found that the absolute values of $v_1$ at $E_{int}/A=120$MeV are smaller than those at lower $E_{int}/A$, i.e. $60$MeV and $90$MeV. The more detailed information of $E_{int}/A$ dependence of $v_1$  at different centralities can be found in Fig.~\ref{fv1CentrEin}(d-f). Because the attractive mean field  plays  a smaller role on nucleons with the increasing of beam energy $E_{int}/A$,  it will transfer the effect on photons' flow then the absolute values of $v_1$ for direct photons and thermal photons  have a fall tendency as shown by nucleons. For thermal photons, the $v_1$ in the centrality of $0\%-20\%$ is much more sensitive to $E_{int}/A$, and  all their values of $v_1$ tend to be a constant when the incident energy  $E_{int}/A$ is larger than $70$MeV.

\begin{figure}
    \includegraphics[width=8.6cm]{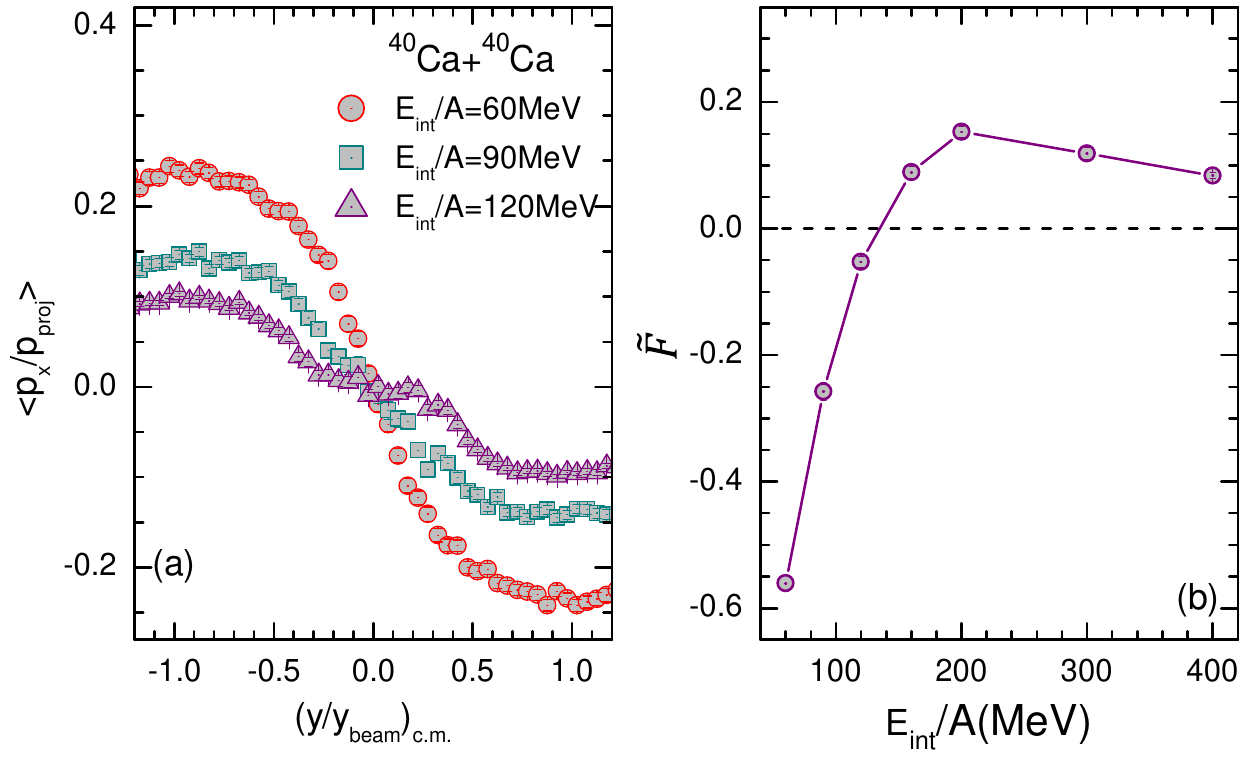}
    \caption{Scale-invariant momentum vs scale-invariant rapidity for mini-bias  $^{40}$Ca+$^{40}$Ca collisions at $E_{int}/A$ = 60, 90 and 120 MeV (a) as well as the flow parameter as a function of incident energy (b). }
    \label{Fflow}
\end{figure}

We noted that a previous study by Bonasera and Csenai suggested that the scaling analysis of transverse flow for charged particles can bring information on change of the reaction mechanism \cite{Bona_PRL}. We checked such an analysis in our Ca + Ca system. Fig.~\ref{Fflow} shows the scale-invariant momentum vs scale-invariant rapidity for mini-bias $^{40}$Ca+$^{40}$Ca collisions at $E_{int}/A$ = 60, 90 and 120 MeV as well as the flow parameter $\tilde{F}$ as a function of incident energy, where  the scale-invariant momentum or rapidity is defined by their values normalized by c.m. momentum ($p_{proj}$) or rapidity ($y_{beam}$) of  projectile, and 
 $\tilde{F}\equiv\frac{d\langle p_x/p_{proj}\rangle}{d((y/y_{beam})c.m.)}|_{(y/y_{beam})c.m.=0}$. Here  the accumulated particles include the ones with charge number $Z$ =1 and 2.   It is seen that there exists a drastic change of the flow sign around 140 MeV/nucleon, indicating  a transition occurs there from the  attractive dominated interaction  to repulsive dominated interaction.

\subsection{Elliptic flows of hard photons and free nucleons}
\begin{figure}
    \includegraphics[width=8.6cm]{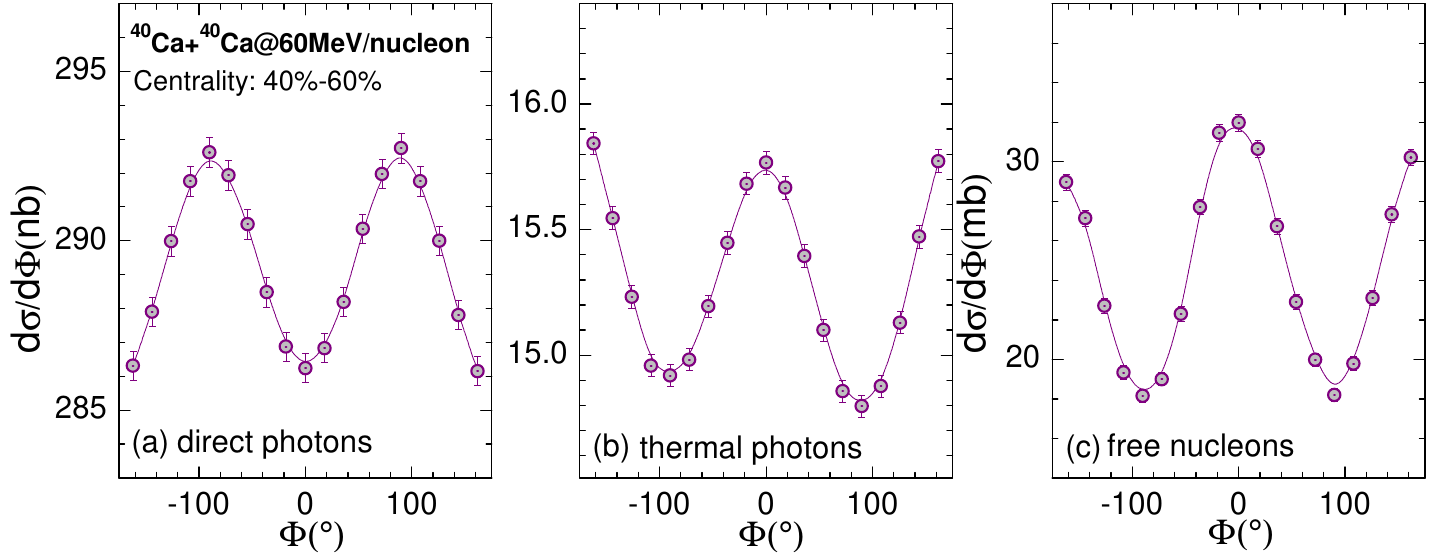}
    \caption{Azimuthal distribution of (a) direct photons, (b) thermal photons and (c) free nucleons for the reaction of $^{40}$Ca+$^{40}$Ca$@60$MeV/nucleon in the centrality of $40\%-60\%$, respectively.}
    \label{fwphi}
\end{figure}

In order to check whether thermal photons have the behavior of elliptic flow $v_2$, Fig.~\ref{fwphi} shows the azimuthal distribution of direct photons, thermal photons and free nucleons for $^{40}$Ca+$^{40}$Ca collision at $60$MeV/nucleon in the centrality of 40\%-60\%, respectively. It is seen that the direct photons prefer to be emitted from out-of-plane while both thermal photons and free nucleons demonstrate in-plane emission enhancement. The phenomenon of thermal photon emission exhibiting a weak collective behavior like direct photons and free nucleons confirms that it also exists azimuthal anisotropic. Therefore, according to the definition of $v_2$ in Eq.~(\ref{v2flow}), one can get the information on elliptic flow $v_2$ for hard photons as well as free nucleons, respectively.

\begin{figure}
    \includegraphics[width=8.6cm]{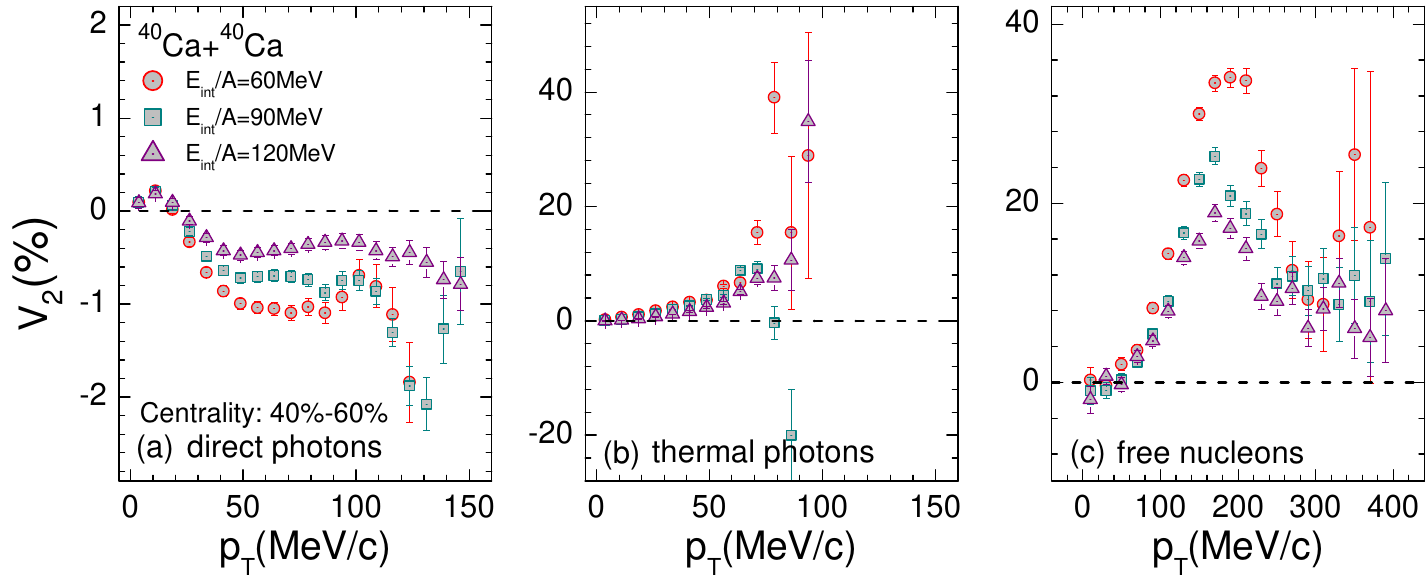}
    \caption{Elliptic flows of (a) direct photons, (b) thermal photons and (c) free nucleons as a function of transverse momentum for $^{40}$Ca+$^{40}$Ca collisions at $60$MeV/nucleon (circles), $90$MeV/nucleon (squares) and $120$MeV/nucleon (triangles) in the centrality of $40\%-60\%$, respectively.}
    \label{fv2pT}
\end{figure}

Fig.~\ref{fv2pT} describes the elliptic flows $v_2$ of direct photons, thermal photons and free nucleons as a function of transverse momentum $p_T$ for $^{40}$Ca+$^{40}$Ca collisions at $60$, $90$and $120$MeV/nucleon in the centrality of $40\%-60\%$, respectively. Similar to the directed flow parameter $v_1$, the values of $v_2$ for direct photons also have the opposite sign with those of thermal photons and free nucleons, which further reflects a different preferential transverse emission in the direction of in-plane or out-of-plane. Moreover, the absolute values of  $v_2$ for all of them have a rise tendency at lower $p_T$. But at higher $p_T$, the $v_2$ of direct photons gradually tends to be saturated while there is a fall tendency for free nucleons. Additionally, it should be noticed that the absolute values of $v_2$ for direct photons are much smaller than that of thermal photons and free nucleons. Meanwhile, we find that the incident energy $E_{int}/A$ plays a reduction effect role for elliptic flows parameter $v_2$ .

 \begin{figure}
    \includegraphics[width=8.6cm]{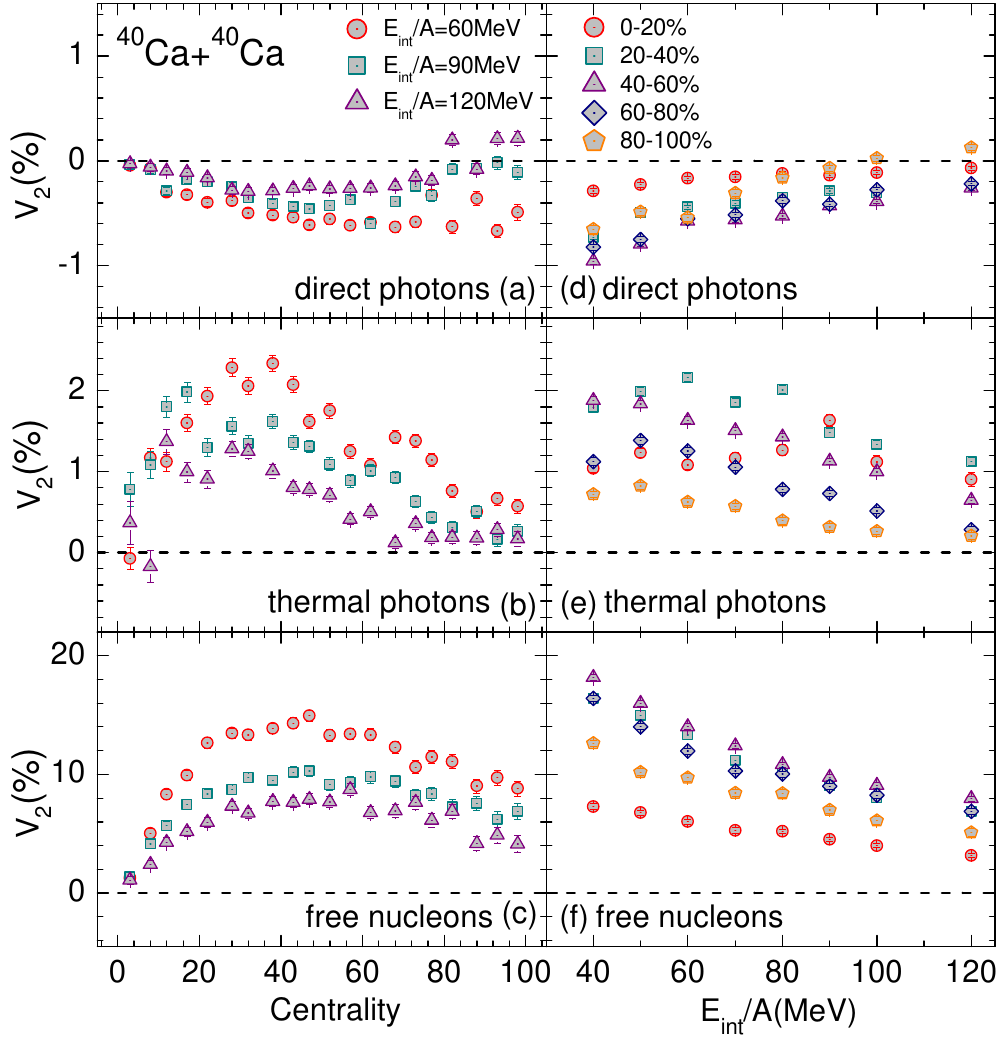}
    \caption{Elliptic flows of direct photons (a and d), thermal photons (b and e) and free nucleons (c and f) as a function of centrality (a-c) and incident energy (d-f) for $^{40}$Ca+$^{40}$Ca collisions, respectively.}
    \label{fv2CentrEin}
\end{figure}

The collision centrality and incident energy dependences of the elliptic flows $v_2$ for direct photons, thermal photons and free nucleons are also investigated in the reaction of $^{40}$Ca+$^{40}$Ca, which are plotted in Fig.~\ref{fv2CentrEin}. All of the absolute values of $v_2$ with collision centrality, i.e. from the central collision to the peripheral collision, have a rise and fall tendency. It can also be found that the behavior of elliptic flows is smaller at $E_{int}/A=120$MeV than those at $60$MeV and $90$MeV, which is similar to the behavior of directed flow $v_1$. More detailed information about the incident energy dependence of the elliptic flows $v_2$ is obtained from Fig.\ref{fv2CentrEin}(d-f). It is well established that the mean field effect as an important part in intermediate-energy heavy-ion collisions tends to reduce with the increase of $E_{int}/A$. As a result, we find that the most absolute values of $v_2$ for direct photons, thermal photons and free nucleons at different centrality range gradually decrease with increasing $E_{int}/A$.  However, there is an interesting phenomenon that the $v_2$ of thermal photons in central collision $\%0-20\%$ has a small fluctuation around a constant $1.1\%$.  What's more, the absolute values of $v_2$ for direct photons and thermal photons  are much smaller than free nucleons by an order of magnitude which reveals the weak effect of azimuthal asymmetry.

\subsection{Multiplicities of hard photons and free nucleons and their correlation}

\begin{figure}
    \includegraphics[width=8.6cm]{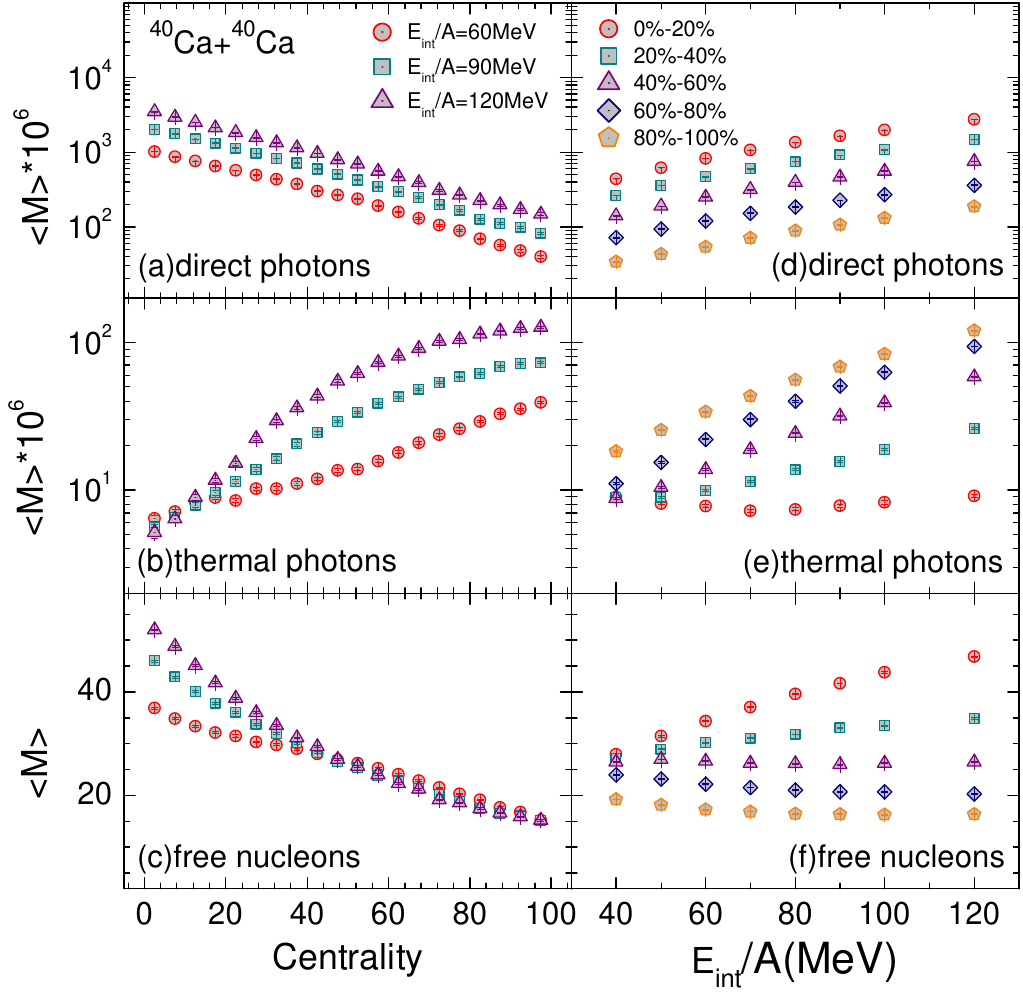}
    \caption{Multiplicities of direct photons (a and d), thermal photons (b and e) and free nucleons (c and f) as a function of centrality (a-c) and incident energy (d-f) for $^{40}$Ca+$^{40}$Ca collisions, respectively.}
    \label{fmult}
\end{figure}

In this section, we continue to study the properties of hard photons and free nucleons by focusing on their multiplicity and their multiplicity correlation in intermediate-energy heavy-ion collisions. Fig.~\ref{fmult} shows the collision centrality and incident energy $E_{int}/A$ dependences of the  multiplicities of direct photons, thermal photons and free nucleons in the reaction of $^{40}$Ca+$^{40}$Ca, respectively. Fig.~\ref{fmult} (a-c) show  that for direct photon and free nucleon, both of their multiplicities rapidly decrease from central collisions to peripheral collisions. However, the multiplicity of thermal photons has a rise tendency and then tends to a saturate state at the higher incident energy $E_{int}/A$, for example at $E_{int}/A=120$MeV. The explanation is that the less free nucleon emission in peripheral collisions means that a larger amount of nucleons are bounded in fragments which enhances the opportunity of individual $n-p$ collision at the later stage of reaction to some extent. Consequently, more thermal photons are emitted from peripheral collisions. On the other side, the properties of the multiplicity versus $E_{int}/A$ at different centrality range can be learnt  from Fig.~\ref{fmult} (d-f).  The multiplicities of direct photons monotonously increase with $E_{int}/A$ for different centralities. But for thermal photons at $0\%-20\%$ centrality, their multiplicities have a decreasing tendency and then has a fluctuation with $E_{int}/A$, which indicates of a fierce competition between thermal photon emission and multi-fragment emission. Multi-fragment production can suppress the emission of thermal photons by reducing the chance of $n-p$ collision, while a higher $E_{int}/A$ produces more light clusters and free nucleons rather than intermediate mass fragment and then contributes to a little more thermal photon production. As for free nucleon, the multiplicity dependence on $E_{int}/A$ has a decreasing tendency with the system from central to peripheral collisions.

In order to understand the correlation between hard photon emission and free nucleon emission, a multiplicity correlation function is constructed as
\begin{equation}
\label{multicorr}
1+ R_{\gamma-n} = \frac{\langle M_{\gamma}\cdot M_{n}\rangle}{\langle M_{\gamma}\rangle\langle M_{n}\rangle},
\end{equation}
where $\langle M_{\gamma}\rangle$ and $\langle M_{n}\rangle $ describe the average multiplicity of hard photons and free nucleons over all events. Note that $M_{\gamma}$ and $M_{n}$ in the term of $\langle M_{\gamma}\cdot M_{n}\rangle$ denote the multiplicities of hard photons and free nucleons produced from the same event. Base on this function, if the correlation factor is equal $1.0$, it manifests that hard photon emission is independent of free nucleon emission. Conversely, the correlation factor is less than $1.0$ which indicates there is an anti-correlation between them, and the correlation factor is larger than 1.0 corresponding a positive correlation between them.

\begin{figure}
    \includegraphics[width=8.6cm]{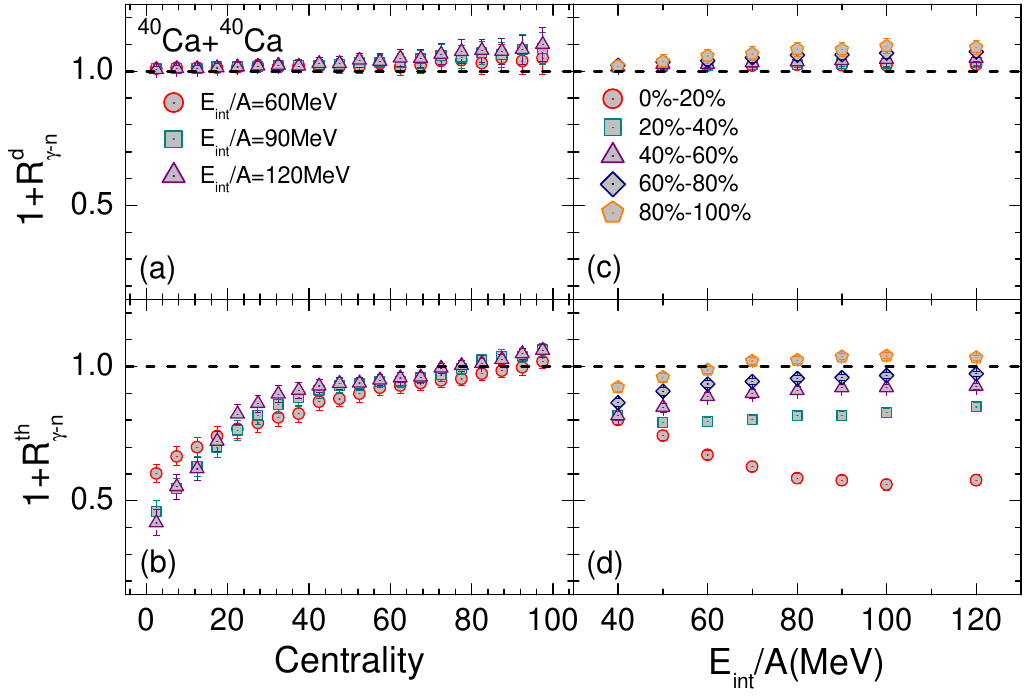}
    \caption{Multiplicity correlation between direct photons (a, c) or thermal photons (b, d) and free nucleons as a function of centrality (a-b) and incident energy (c-d) for $^{40}$Ca+$^{40}$Ca collisions, respectively.}
    \label{fcorrmult}
\end{figure}

Fig.~\ref{fcorrmult}(a) and (b) plot the multiplicity correlation between hard photons and free nucleons as a function of collision centrality for $^{40}$Ca+$^{40}$Ca collisions at $60$, $90$ and $120$MeV/nucleon, respectively. For direct photons, the correlation factor $1+R_{\gamma-n}$ is larger than $1.0$ and it gradually enhances with increasing centrality. However, the correlation parameter distribution of thermal photons with centrality is completely different from that of direct photons, i.e. the parameter $|R_{\gamma-n}|$ between thermal photons and free nucleons tends to reduce from central to peripheral collision, and thermal photon emission is negatively correlated with free nucleons emission. Meanwhile, we find the $|R_{\gamma-n}|$ for direct photons is much smaller than that of thermal photons, especially in central collisions. It confirms that much more free nucleons are emitted in central collision and then it inhibits the thermal photon production at the later stage of reaction.

Fig.~\ref{fcorrmult}(c) and (d) display the multiplicity correlation between hard photons and free nucleons versus incident energy $E_{int}/A$ for different collision centralities. It shows that the correlation factor $1+R_{\gamma-n}$ of direct photons is just a little larger than $1.0$, which also indicates there is a weak positive multiplicity correlation between direct photon emission and free nucleon emission. For thermal photons emitted from peripheral collisions, the correlation factor has a small fluctuation with $E_{int}/A$. But for central collisions, i.e. at $0\%-20\%$ centrality, the $|R_{\gamma-n}|$ has a rise tendency and then tend to be a constant when $E_{int}/A$ is larger than $70$MeV, which is similar to the $v_1$ distribution with incident energy.

\section{Summary}
\label{summary}
In summary, the incoherent neutron-proton bremsstrahlung photon production channel was embedded into the IQMD model in order to investigate high energy photon properties.  In this work, we mainly focused on the investigation of anisotropic flow and multiplicity of hard photons and free nucleons, which are produced in intermediate-energy heavy-ion collisions.  Firstly, the distribution of the average in-plane transverse momentums $\langle p_{x}\rangle$ versus  the rapidity of  particles in the center-of-mass confirmed that thermal photons have a directed flow behavior. Then a systematic study of the directed flow $v_1$ for direct photons, thermal photons and free nucleons were performed, including the dependences of transverse momentum $p_T$, collision centrality and incident energy $E_{int}/A$. The results showed that direct photons mainly stem from the repulsive mean field. But both thermal photons and free nucleons were dominated by the attractive mean field. Secondly, the azimuthal distribution of hard photons and free nucleons also demonstrated that thermal photons exists an elliptic flow behavior. We found that direct photons have a different preferential transverse emission from thermal photons and free nucleons. The former tends to be emitted from out-of-plane and the later prefers to from in-plane. Finally, the collision centrality and incident energy dependences for hard photons and free nucleons are also discussed.  In order to further understand the properties of hard photons and free nucleons, a multiplicity correlation is constructed. It is found that there is a weak multiplicity correlation between direct photons and free nucleons comparing with that between thermal photons and free nucleons. Meanwhile, a positive correlation is observed between direct photon emission and free nucleons emission, and an anti-correlation correlation is found between thermal photon emission and free nucleons emission. In addition, for thermal photons, its directed flows $v_1$ versus  $E_{int}/A$ have a similar tendency with the multiplicity correlation parameter $R_{\gamma-n}$ in central collisions, i.e. their absolute values  have a rise tendency and tend to be a constant when $E_{int}/A$ is larger than $70$MeV.

\begin{acknowledgments}

This work was supported in part by the National Natural Science Foundation of China under Contracts Nos. 11890714, 11421505, 11925502, 11975091, 11305239, 11220101005 11961141003 and 11935001, the Strategic Priority Research Program of the Chinese Academy of Sciences under Grant No. XDB34030200  and XDB16, % the Major State Basic Research Development Program in China under Contract No. 2014CB845401,
the Youth Innovation Promotion Association CAS under Grant No. 2017309.
%, the Program for Science and Technology Innovation Talents in Universities of Henan Province under Grant No. 13HASTIT046, Natural and Science Foundation in Henan Province under Grant No. 162300410179.

\end{acknowledgments}

\end{CJK*}
\end{document}